\documentclass[twocolumn,superscriptaddress,amsmath,amssymb,floatfix,aps,notitlepage,nokeys,prl,longbibliography,nofootinbib]{revtex4-1}
\usepackage{mathrsfs, hyperref}
\usepackage{amssymb, amsbsy, amsmath, latexsym, dsfont, array, layout, graphics,mathrsfs,braket,amsfonts,amsthm,
  amssymb,graphicx,subfigure,dcolumn, youngtab,color,bm,mathtools,braket,verbatim,url,cleveref,natbib,hypernat,extarrows,inputenc,multirow}

\usepackage[mathlines]{lineno}

\begin{document}
\preprint{APS/123-QED}

\title{Drops on architected elastic substrates: A repertoire of regimes at the turn of a knob} 

\author{Sergio Santos}
\affiliation{%
Department of Civil, Environmental, and Geo- Engineering,
University of Minnesota, Minneapolis, MN 55455, US}
\author{Zakari Kujala}
\affiliation{%
Department of Mechanical Engineering,
University of Minnesota, Minneapolis, MN 55455, US}
\author{Sungyon Lee}
\affiliation{%
Department of Mechanical Engineering,
University of Minnesota, Minneapolis, MN 55455, US}
\author{Stefano Gonella}%
\email{sgonella@umn.edu}
\affiliation{%
Department of Civil, Environmental, and Geo- Engineering,
University of Minnesota, Minneapolis, MN 55455, US}%
\date{\today}

\begin{abstract}

Drops on a vibrating substrate can experience a variety of motion regimes, including directional motion and climbing. The key ingredient to elicit these regimes is simultaneously activating the in-plane and out-of-plane degrees of freedom of the substrate with the proper phase difference. This is typically achieved by using a rigid substrate and two independent actuators. 
However, this framework is unable to establish different motion conditions in different regions of the substrate, achieving spatial variability and selectivity, since this would violate the rigid-body assumption and require a proliferation of actuation channels. Challenging this paradigm, we leverage the inherent elasticity of the substrate to provide the modal and spatial diversity required to establish the desired regimes. To this end, we design deformable substrates exhibiting a rich landscape of deformation modes, and we exploit their multi-modal response to switch between drop motion regimes and select desired spatial patterns, using the excitation frequency as our tuning parameter.

\end{abstract}

\maketitle
      
The ability to control confined amounts of liquids has applications in microfluidics~\cite{shesto2004, Seemann2012, Stone2004, Squires2005}, coating and printing processes~\cite{Calvert2001, Modak2020}, particle suspension~\cite{Whitehill2010}, electric generators~\cite{Mugele2005, Li2022}, and biological processes~\cite{Blossey2002, Yu2019, Zwicker2025, Zeng2023}. Of particular interest is the problem of 
controlling the movement of sessile drops placed on a surface. Here, to promote drop motion, one must overcome the pinning forces (comprehensively described by contact angle hysteresis (CAH)) generated by physical or chemical 
interactions with the surface~\cite{Costalonga2020, Joanny1984, Gennes1985, Eral2013}. 
Drop motion can be controlled through a variety of strategies, including surface chemical treatment~\cite{Huang2014, Varagnolo2014, Lin2018} or structural modifications~\cite{Buguin2002, Sandre1999, Bintein2019}, and mechanical vibrations of the substrates~\cite{Daniel2002, Daniel2004, Daniel2005, Whitehill2010, Sartori2015, Noblin2009, Brunet2007, Costalonga2020, Dong2006, Celestini2006, Brunet2009, John2010, Deegan2020}. 

In the context of vibration-induced drop control, achieving \textit{net} drop motion requires breaking the problem symmetry, which is accomplished by overcoming CAH in a predominant biased direction over each excitation cycle. Intriguing motion scenarios have been realized following this concept, including drops that can overcome gravity and climb on inclined surfaces~\cite{Brunet2007}.
%
Canonically, drops are moved under vibration by rigidly translating the entire substrate~\cite{Brunet2007,Brunet2009,Costalonga2020,Noblin2009,Sartori2015}. This is attained by simultaneously controlling the in-plane (IP) and out-of-plane (OOP) degrees of freedom of the plate, either by prescribing an excitation along a given direction or by using independent, dedicated actuators for the IP and OOP components~\cite{Noblin2009}.
Either way, the relative amplitude and phase between IP and OOP are the key parameters that dictate the direction of motion. 
A major drawback of working with rigid substrates is a severe lack of spatial tuning and local control: since every point on the plate experiences the same kinematics, drops deposited on different regions inevitably experience the same motion conditions. 
Our goal is to inject an additional \textit{spatial selectivity} into the substrate, thus promoting 
its ability to subject different subsets of drops to different motion regimes, according to a variety of spatial patterns. 

One avenue to achieve spatial selectivity is to work with \textit{elastic} substrates, where a spatial diversity of the mechanical response stems naturally from the internal deformability that is inherent to all elastic media. 
Moreover, an additional diversity of deformation profiles is encoded in the spectrum of mode shapes that elastic structures can experience even under a single, fixed source of excitation.
This consideration inspires the development of a parsimonious experimental framework that relies exclusively on the elastic response of the substrates to achieve all the desired drop motion control patterns 
without unnecessary proliferation of actuation sources. 
The potential of elastic substrates for spatially-selective control was initially explored in~\cite{Charara-et-al_Metaplate-Drops_PRSA_2025} using thin metaplates as vibrating substrates. However, while this approach was successful at 
programming depinning and the onset of sliding, other useful regimes, such as net motion on a horizontal surface~\cite{Noblin2009,Costalonga2020} or climbing~\cite{Brunet2007}, remained unattainable. To realize these conditions, a substrate must experience appreciable IP deformation. Unfortunately, classical plates are significantly stiffer IP than OOP, making their IP response negligible. To fully unleash the potential of elastic substrates, it is imperative to design special substrates that display comparable IP and OOP compliance.

\begin{figure}[t!]
\includegraphics[width=\columnwidth]{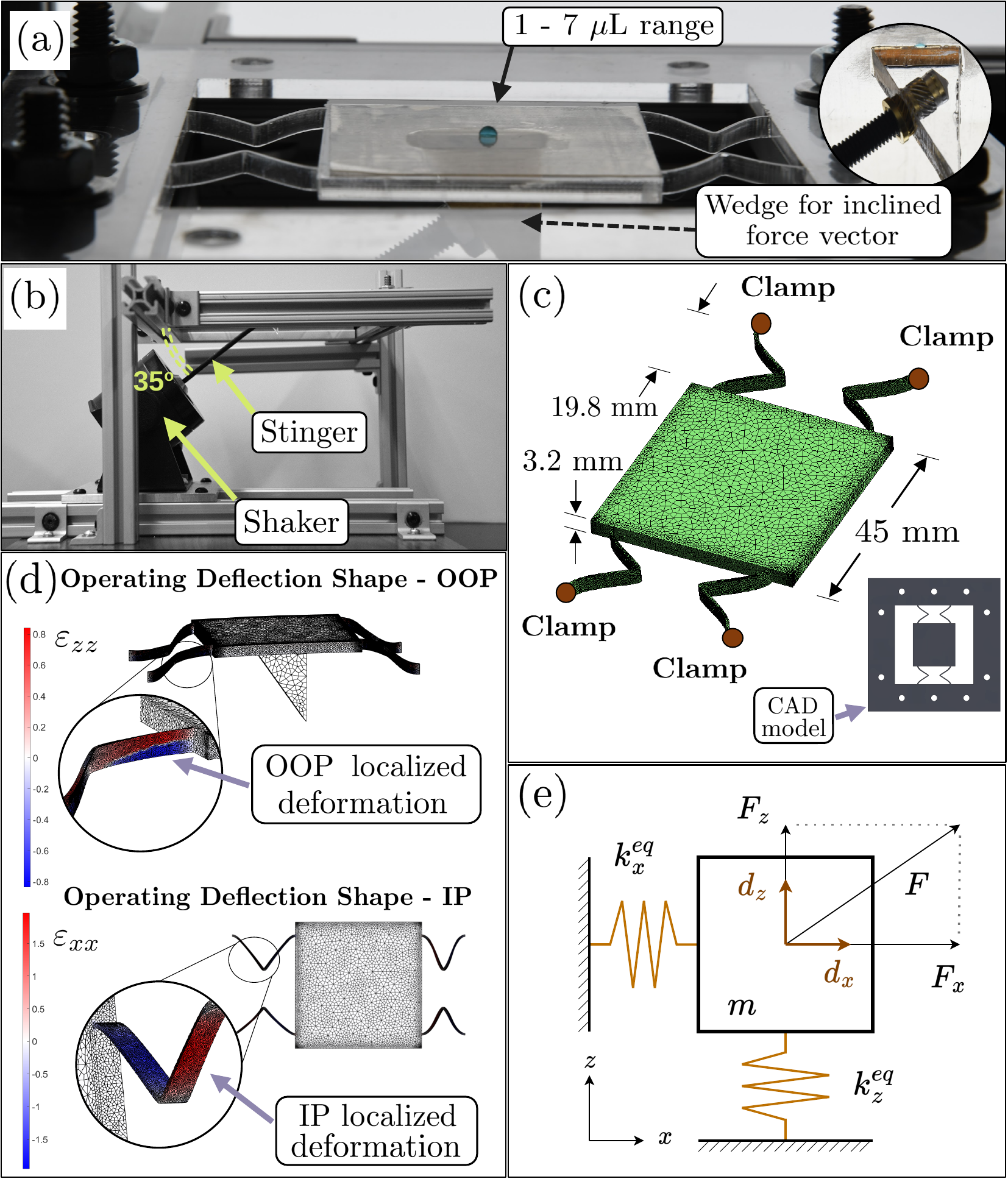}
\caption{\label{fig:panel1}(a) Single-cell prototype, with inclined excitation prescribed through actuation wedge. (b) Experimental setup showing support frame and shaker. (c) Meshed model used in finite element (FE) simulations. (d) Operating deflection shapes for OOP and IP motion, providing evidence of deformation localizing at the ligaments. (e) Equivalent lumped-parameter single-body system with mass $m$ and equivalent springs.
The system is excited by harmonic force $F$ and undergoes IP $d_x$ and OOP $d_z$ displacements.}
\end{figure}

In this letter, we present a strategy to achieve this goal. The inspiration comes from the design philosophy of elastic metamaterials, whose archetypal architecture often consists of a discrete network of masses joined by structural objects that serve as elastic connections. Accordingly, we envision our substrate 
as a discrete set of one or more plates (referred to as \textit{cells}), which can be taken to be locally rigid, supported by structural ligaments designed to localize the 
deformation and maximize the global IP compliance. 
Following this idea, we assemble the 
prototype shown in Fig.~\ref{fig:panel1}(a), made by laser cutting a thin acrylic sheet (Young's modulus $E=3.1 \, \textrm{GPa}$, Poisson ratio $\nu=0.35$, thickness $3.2 \, \textrm{mm}$), and the setup depicted in Fig.~\ref{fig:panel1}(b). 
The cell is a $45 \, \textrm{mm} \times 45 \, \textrm{mm} $ plate connected to a fixed frame by doubly-curved thin ligaments featuring high IP compliance.
An acrylic wedge is attached to the bottom surface to exert a force at a prescribed angle (here $35^{\text{o}}$ from the horizontal) via the stinger of a 
shaker (Br\"{u}el \& Kj\ae r Type 4810 Mini). 
Given the contrast in stiffness between plate and ligaments, we can assume that the deformability 
is concentrated in the ligaments and the plate undergoes rigid-body motion. This assumption is validated via finite element (FE) simulations (model mesh shown in Fig.~\ref{fig:panel1}(c)), which confirm that the IP and OOP response 
is 
indeed localized in the ligaments, see Fig.~\ref{fig:panel1}(d). Accordingly, the ligaments can be functionally interpreted
as a pair of equivalent springs (one for IP, one for OOP) and the system can be conceptually reduced to the lumped-parameter single-body model in Fig.~\ref{fig:panel1}(e), featuring two displacement degrees of freedom (IP $d_x$ and OOP $d_z$). 
This reduced-order description emphasizes the mechanistic role played by the prototype parts and would qualitatively capture the low-frequency dynamics of the substrate in our excitation band of interest, e.g., the resonance modes in Fig.~\ref{fig:panel1}(d). Nevertheless, for additional quantitative accuracy and generality, we will use the FE model for all calculations in the remainder of the Letter.
%

We deposit on the cell surface water drops with volume in the 1--7 $\mu \text{L}$ range  (see Fig. 1(a)). For this volume interval, using the semianalytical sessile drop eigenfrequency criteria in ~\cite{Celestini2006, Sartori2015, Toth2011, Oliver1977}, we estimate the following resonant frequency ranges: for rocking mode 38--92 Hz and for pumping mode 76--184 Hz (see Fig. 1 in supplemental material (SM)). In our experiments, we carefully target the interval between these two modes, which ensures that drops in our prescribed volume range 
will be in their ideal frequency range for controlled motion, as discussed in~\cite{Noblin2009}. 
To this end, we deliberately design the structural features of the substrate to yield a rich and well-defined modal response 
in this interval, as we will substantiate below. 
To make the surface slightly hydrophobic to promote drop motion~\cite{Costalonga2020}, 
we apply a low-friction tape with measured equilibrium contact angle $94^o \pm 3^o$, advancing angle $92^o \pm 4^o$, and receding angle $73^o \pm 3^o$. 

\begin{figure}[t!]
\includegraphics[width=\columnwidth]{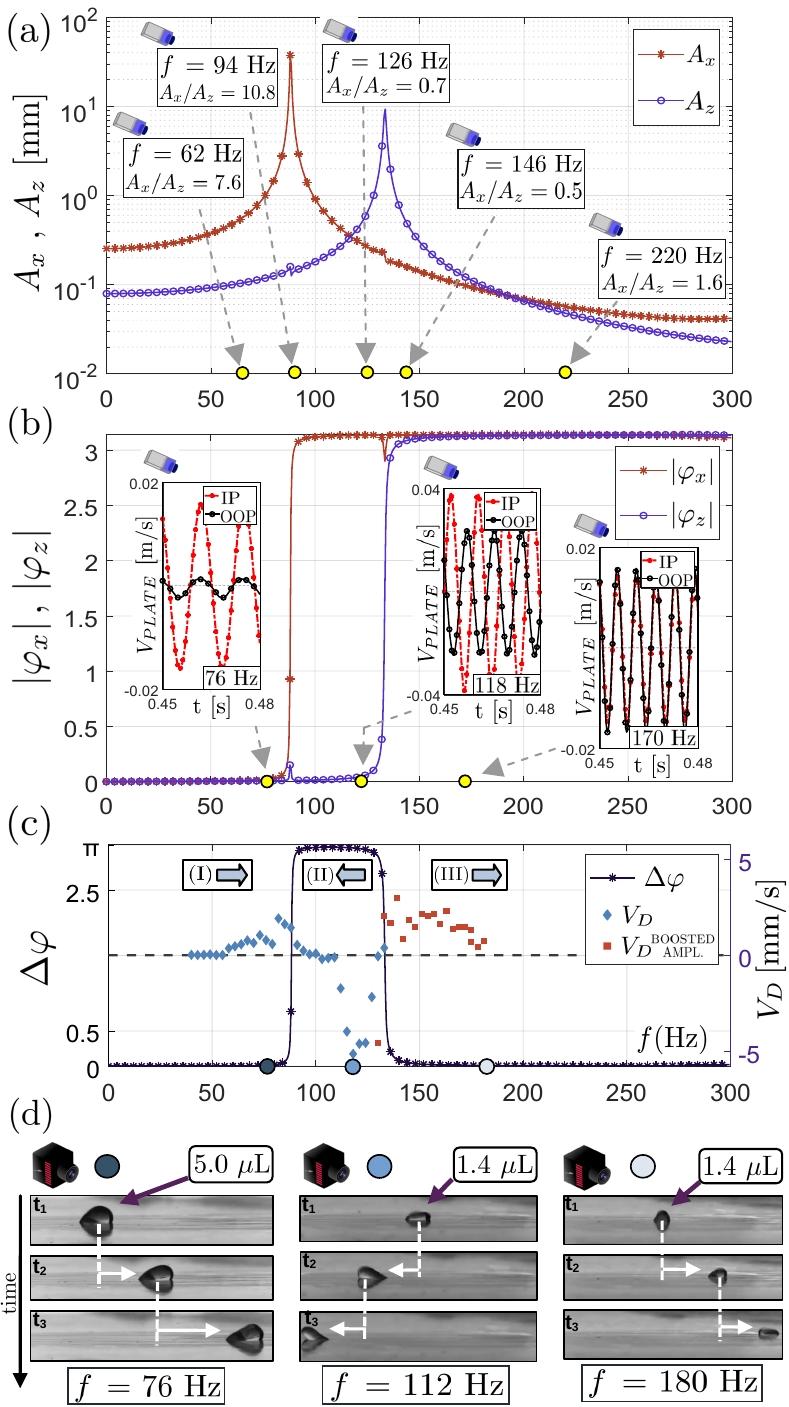}
\caption{\label{fig:panel2}(a) FE-computed spectral amplitudes $A_x$ and $A_z$ for $d_x$ and $d_z$ displacements, respectively. The laser heads icons denote laser vibrometer readings at sampled frequencies. (b) Displacement-force phases $|\varphi_x|$ and $|\varphi_z|$, featuring jumps at natural frequencies. The insets show substrate velocity time histories acquired from laser measurements, confirming phase switches at resonances and antiresonances. (c) Relative phase metric $\Delta \varphi = ||\varphi_x| - |\varphi_z||$ responsible for selection of drop motion direction, plotted vs. $f$ and experimentally measured drop velocities $V_D \, \text{[mm/s]}$ at sampled frequencies. (d) High-speed camera captures of drop motion at frequencies $76 \, \textrm{Hz}$, $112 \, \textrm{Hz}$, and $180 \, \textrm{Hz}$ confirming predictions of drop directions.}
\end{figure}

We characterize the dynamics of the cell in Fig.~\ref{fig:panel1} using finite element simulations under a harmonic point excitation mimicking the stinger inclination. 
The two curves in Fig.~\ref{fig:panel2}(a) show the spectral amplitudes $A_x$ (IP) and $A_z$ (OOP) computed at the central node of the mesh as a function of frequency $f$. We observe two distinct natural frequencies, one at $\approx 90 \, \textrm{Hz}$, associated with IP motion, the other at $\approx 135 \, \textrm{Hz}$, associated with OOP, which subdivide the domain into three frequency intervals. We corroborate this trend experimentally with a few laser-vibrometer measurements (insets denoted by laser head icons) which by and large confirm the ratios between IP and OOP displacements predicted by FE at the sampled frequencies. Here we expect that, for each degree of freedom, every resonance peak induces an abrupt shift by $\pi$ in the phase between the displacement and the force that causes it. Indeed, in Fig.~\ref{fig:panel2}(b) the phase components $\varphi_x$ (IP) and $\varphi_z$ (OOP) jump at their respective resonance frequencies, again marking the three characteristic frequency regimes mentioned above. Note that a small amount of viscoelastic damping is introduced in the FE model, which smoothens the jumps in the phase curve. The insets in Fig.~\ref{fig:panel2}(b) show laser-acquired plate velocity time histories at three representative frequencies (selected within our three regions of interest), which experimentally confirm the expected phase switches. 

The quantities plotted in Fig.~\ref{fig:panel2}(b) are phase differences between each displacement component ($d_x$ or $d_z$) and their corresponding force component ($F_x$ or $F_z$). Since the force is prescribed by a single actuator, $F_x$ and $F_z$ are naturally in phase with respect to each other. 
In light of this, 
we can 
define a relative phase metric $\Delta \varphi 
\overset{\Delta}{=} ||\varphi_x| - |\varphi_z||$, plotted in Fig.~\ref{fig:panel2}(c),  
to capture, with a single curve, the effective difference in phase between the displacement components $d_x$ and $d_z$.  
Since it has been shown 
that the key ingredient to control (and possibly invert) the direction of drop motion on a horizontal surface is precisely the ability to control the phase between IP and OOP displacements, a simple inspection of the frequency regions emerging in the $\Delta \varphi$ curve is sufficient to predict all attainable motion regimes. 
Specifically, we appreciate that regions I (low frequencies) and III (high frequencies), where $\Delta \varphi \approx 0$, feature in-phase motion between IP and OOP displacements; in contrast, in region II between resonances (mid frequencies), where $\Delta \varphi \approx \pi$, we achieve opposition of phase. 
In Fig.~\ref{fig:panel2}(c), the boxed arrows indicate the three drop motion directions (right (R), left (L), and again right (R)) predicted by theory for the three phase regimes, respectively. 
Let us emphasize an important departure from the canonical paradigm involving a rigid substrate and distinct actuators for IP and OOP (see, for instance, the benchmark results in~\cite{Noblin2009}), where inversion of the direction of drop motion is obtained by controlling the relative phase of the two actuators. Here, we achieve the same result with a single actuator, by only changing the frequency and relying on the inherent structural dynamics of the cells to provide the necessary amplitude and phase conditions.  

In Fig.~\ref{fig:panel2}(d), we substantiate this claim 
experimentally through three sets of snapshots from high-speed videos for each regime. The cases correspond to frequencies $76 \, \textrm{Hz}$, $112 \, \textrm{Hz}$, and $180 \, \textrm{Hz}$, marked by color-coded dots on the $f$ axis in Fig.~\ref{fig:panel2}(c). The snapshots over three time instants document drop motion directions consistent with the predictions. While we only show snapshots for one representative frequency per region, we measure drop velocities for a host of frequencies.
%
%
We superimpose the measured drop velocities on the $\Delta \varphi$ 
curve in Fig.~\ref{fig:panel2}(c). In order to ensure that drops receive enough input energy as the IP response naturally decreases at high frequencies, we increase the amplitude of the excitation prescribed to the system (labeled ``boosted amplitude" in legend) above $\approx$ 140 $\textrm{Hz}$ (details in Table 1 in SM). 
The measured velocities 
confirm our predictions, with right- and left- moving drops appearing in the in-phase and out-of-phase regime, respectively.

\begin{figure}[t!]
\includegraphics[width=\columnwidth]{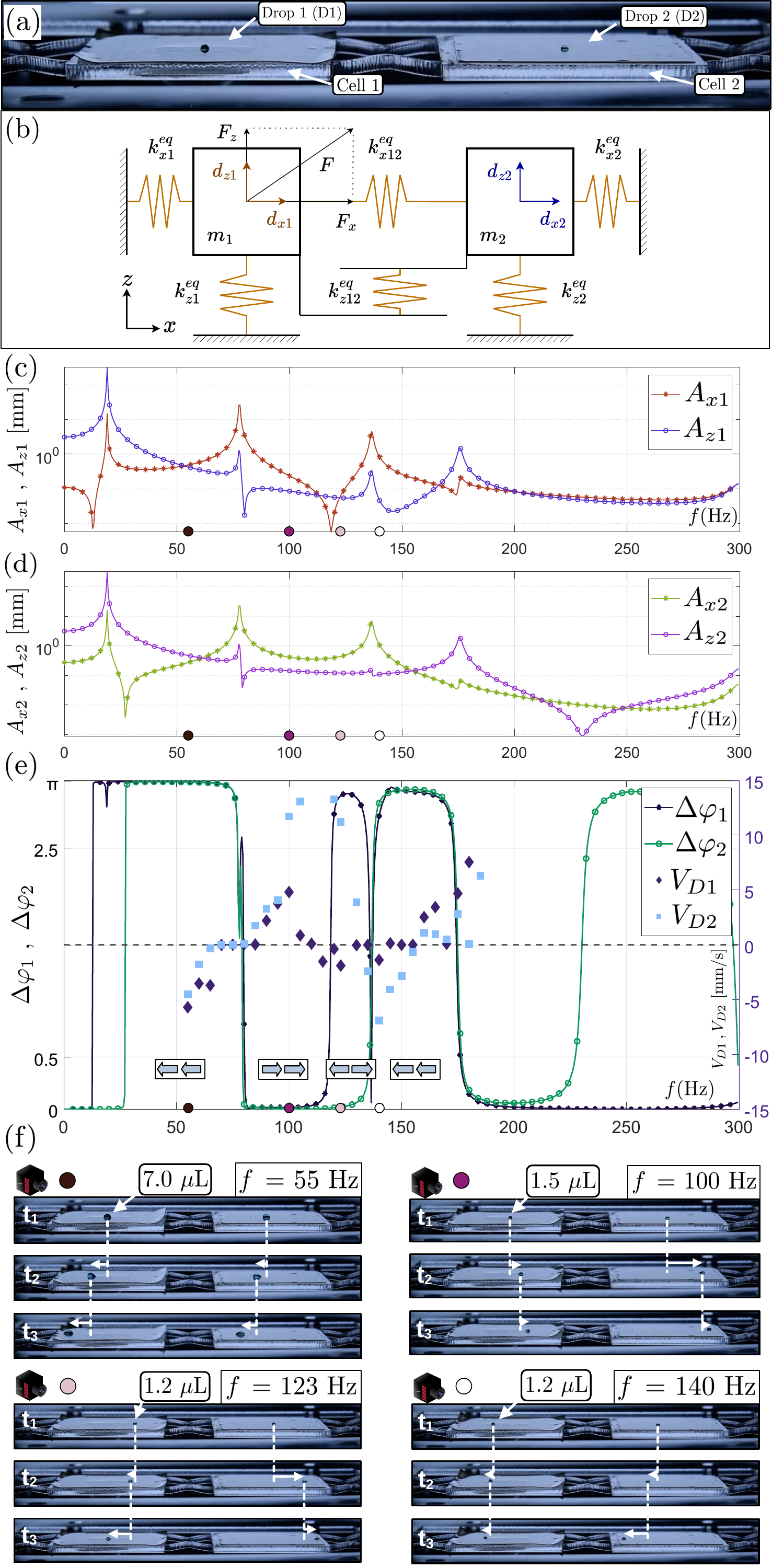}
\caption{\label{fig:panel3}(a) Two-cell prototype. (b) Equivalent lumped-parameter model. 
A harmonic force is applied to cell 1. (c-d) Spectral amplitudes $A_{x1}$ (IP) and $A_{z1}$ (OOP) for cell 1 and $A_{x2}$ (IP) and $A_{z2}$ (OOP) for cell 2, from FE. (e) Relative phases $\Delta \varphi_1$ and $\Delta \varphi_2$ determining the R vs. L motion of their respective drops. Drop velocities experimentally acquired at sampled frequencies are overlaid to the plot, showing good correlation with the $\Delta \varphi$ curves. (f) High-speed camera snapshots revealing activation of LL, RR, LR (divergent) and again LL drop motion at $55 \, \textrm{Hz}$, $95 \, \textrm{Hz}$, $123 \, \textrm{Hz}$ and $140 \, \textrm{Hz}$.}
\end{figure}

So far, we have shown that we can select the direction for drop motion on a single vibrating substrate by using the excitation frequency as the tuning parameter. The next natural question is: does this capability hold for more complex structures? If so, what additional drop motion regimes can be elicited? One way to add geometric complexity is by adding masses (and degrees of freedom) to the vibrating system. To this end, we augment the system, doubling the cells and connecting them through a pair of ligaments identical to those that constrain the cells to ground, eventually obtaining the two-cell prototype shown in Fig.~\ref{fig:panel3}(a). The goal is to activate the motion of both cells through a single excitation applied to cell 1, relying exclusively on the elasticity of the system to transfer the excitation to cell 2. This augmented system can be interpreted as the two-body lumped-parameter model shown in Fig.~\ref{fig:panel3}(b), which features 4 degrees of freedom, one IP and one OOP per cell. FE simulations confirm that the compliance of the cells is negligible and the deformation localizes in the ligaments for the frequencies of interest in our experiments; therefore, the cells behave effectively as rigid bodies connected by three pairs of identical equivalent springs that fully capture the system's elasticity.

To characterize the system, we can extend the analysis for the single cell, accounting for the extra degrees of freedom. 
The results in Figs.~\ref{fig:panel3}(c) and~\ref{fig:panel3}(d) show the spectral amplitudes for cells 1 and 2, respectively. 
The phase is reported in Fig.~\ref{fig:panel3}(e) where, for each cell $i=1,2$, the $\Delta \varphi_i$ curve captures the phase between IP and OOP components. By treating $\Delta \varphi_i$ as a predictor of the direction of drop motion on the $i$ cell, we deduce that we can achieve up to four possible scenarios in the 50--175 Hz interval: (i) both drops move left (LL), (ii) both move right (RR), (iii) the drops move away from each other (LR or divergent pattern), and (iv) the drops move towards each other (RL or convergent pattern). 

In Fig.~\ref{fig:panel3}(f), we show the experimental results for a few drop motion regimes we manage to activate in practice. We monitor the drop velocity time histories captured at selected frequencies falling in four characteristic frequency intervals (indicated by the color-coded dots). Similarly to the single-cell, we adjust the drop size for each frequency interval (details in Table 2 in SM). The snapshots document the establishment of the regimes expected in these intervals: LL, RR, LR and again LL.
These results by and large confirm 
that we can establish a multitude of drop motion scenarios with a single actuator, relying on frequency selection as our tuning parameter. This said, many factors contribute to how easily we can trigger the predicted regimes, and some scenarios are especially challenging to obtain. For example, we can see that the low OOP amplitude $A_{z1}$ predicted by FE simulations between $100 \, \textrm{Hz}$ and $150 \, \textrm{Hz}$ penalizes drop 1 movement in this range. This is confirmed by the low experimental values $V_{D1}$ measured in this region. We also report a small discrepancy between our relative phase $\Delta \varphi$ prediction and the experimentally observed net velocity $V_{D}$, for both drops, above $150 \, \textrm{Hz}$. This suggests some mismatch between FE model and prototype occurring at higher frequencies, with the former predicting a resonance at $\approx$ 175 Hz and the drops velocity indicating one at $\approx$ 160 Hz.
For completeness, in SM, we present an alternative prototype design that 
realizes the fourth scenario (RL) not achieved in the system of Fig.~\ref{fig:panel3}.

\begin{figure}[t!]
\includegraphics[width=\columnwidth]{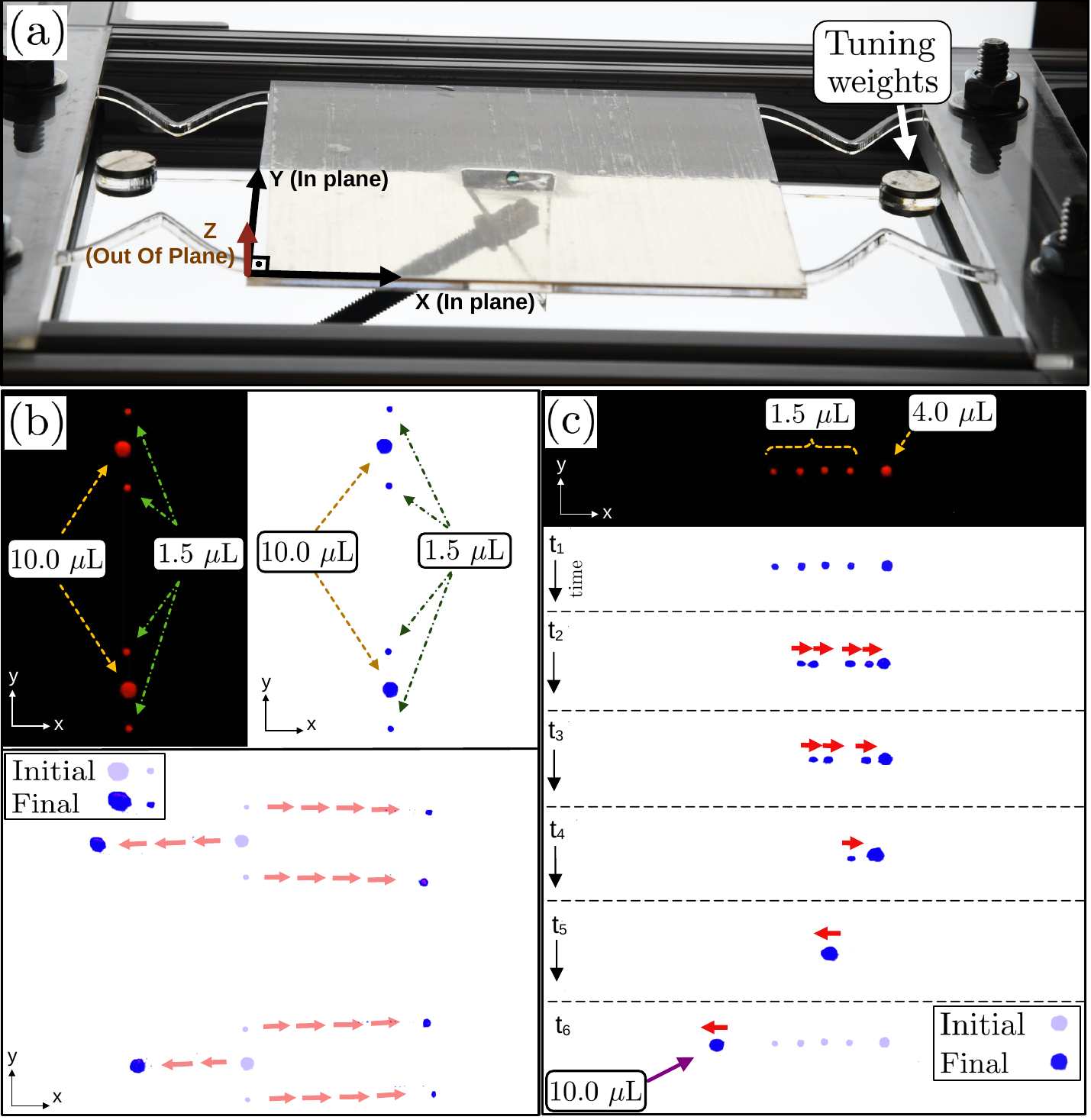}
\caption{\label{fig:panel4}(a) Experiment with modified 
prototype. Tuning weights added to cell in detail. (b) Size-selective filter achieving drop segregation by size. Dark picture shows original acquisition with fluorescent drops; light images are post-processed for enhanced visual clarity. Movement is shown by red arrows. (c) Coalescence, growth and path inversion device, featuring drop motion inversion due to variable size triggered by controlled coalescence of a train of drops.}
\end{figure}

The fact that drop motion depends on both frequency and drop size allows us to create a variety of drop motion logic scenarios even operating with a single plate.
Here we discuss two of them. For this task, 
we need to introduce some structural modifications to the first prototype: 
(i) the cell size is increased to grant the drops more freedom of motion and allow observing pattern formation, and (ii) in order to lower the frequency spectrum, resonators are added to the plate in the form of two symmetric cantilevers with tunable tip masses ((Fig.~\ref{fig:panel4}(a)), details in Fig. 2 in SM). The first scenario realizes a \textit{size-selective drop filter} (Fig.~\ref{fig:panel4}(b)). Four 1.5 $\mu \text{L}$ drops and two 10 $\mu \text{L}$ drops are deposited at the same axial coordinate. As the cell is vibrated, the two groups of drops move in opposite directions, effectively segregating from each other. This follows from the fact that drops of different sizes have different internal resonance dynamics~\cite{Mettu2012, Steen2019, Dong2006, Costalonga2020} and therefore respond differently to the same frequency of excitation. Here, the smaller drops move right, as predicted by the logic of Fig.~\ref{fig:panel1}, assuming that they respond in 
their first resonance range. At the same frequency, larger drops experience a higher mode, and it has been shown~\cite{Dong2006, Costalonga2020} that, when this occurs, the dependence of the drop directionality upon the substrate phase is inverted. This example shows how the motion regimes depend on the resonant dynamics of both the cell and the drop, and their interplay. The second system is a \textit{coalescence, growth and path inversion} device (Fig.~\ref{fig:panel4}(c)). Here, a heterogeneous array of drops 
is deposited along a line. All the drops but the rightmost one are small. As the cell is vibrated, the small drops move to the right, colliding, merging and growing in volume. They ultimately merge with the large drop, acquiring a volume that exceeds the threshold above which the directionality switches. At this stage, the emerging large drop starts moving to the left. This operation can be seen as a \textit{fluidic seeker mission}, where a drop that had previously traveled in a certain direction can be recalled to its starting position by sending a train of smaller drops to seek contact with it.

Ultimately, the 
two-cell prototype represents a first step toward a powerful generalization, where a prototype of $n$ cells (effectively a \textit{metachain}) could generate, in principle, up to $2^n$ combinations of relative drop motions. The problem would eventually morph into a wave propagation problem best described in terms of dispersion, and the appearance of bandgaps in the spectrum could be exploited to suppress drop motion at selected locations. 
On the other hand, we have shown that even the basic single-cell substrate has significant potential for drop control when the availability of frequency-dependent and drop-size-dependent regimes is leveraged in scenarios involving multiple drops activated simultaneously or sequentially. Extending this idea to problems involving large and heterogeneous families of drops promises to further enhance the pattern formation and control capabilities hinted at in Fig.~\ref{fig:panel4}(b). Future efforts will assess the practical feasibility of these concepts.


The authors acknowledge support from the National Science Foundation (NSF grant CMMI-2211890).


%

\clearpage

\section{Supplemental Material}

\subsection{Drop size choice - Drop modes}

\begin{figure}[t!]
\includegraphics[width=\columnwidth]{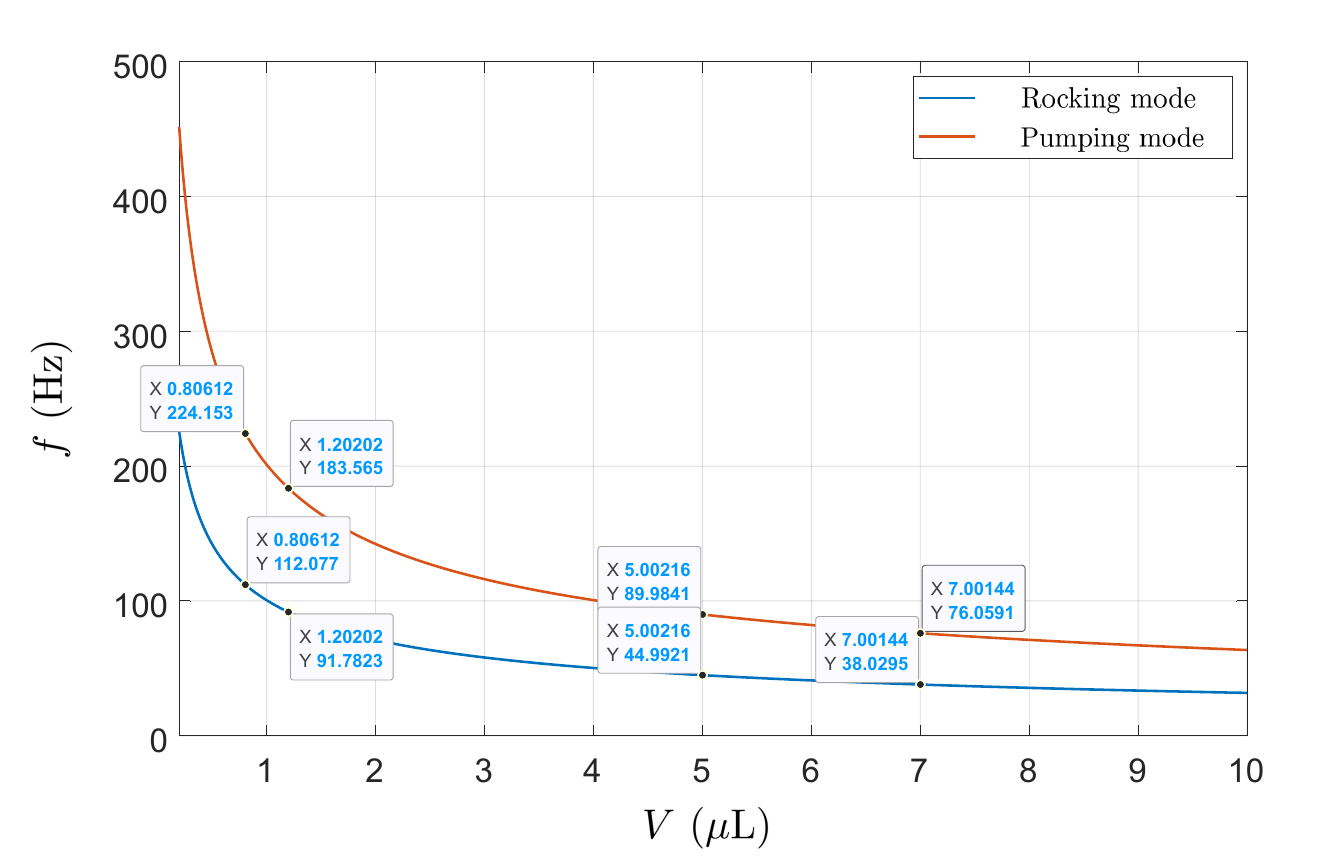}
\caption{\label{fig:dropsize_rockingfrequency} Rocking mode (lower bound) and pumping mode (upper bound) as function of drop volume. Highlight on some drop sizes utilized.}
\end{figure}

Drops display a broad spectrum of mode shapes~\cite{Mettu2012, Steen2019}. As mentioned in the main text, if the changes in $f$ in the sweep are drastic, we may fundamentally alter the modal response of the drops; in order to maintain the drop response in the neighborhood of the same (first) \textit{rocking mode} for all values of $f$, we adjust the drop size from case to case (from 1.2 to 5.0 $\mu \text{L}$), following the semianalytical sessile drop eigenfrequency $f_0$ criteria from \cite{Celestini2006} (which we explicitly write in Eq.~\ref{eq:rockingmode}) along with the drop volume approximation as a spherical cap \cite{Sartori2015, Toth2011, Oliver1977} (depicted in Eq.~\ref{eq:dropcapvolume}), which depends on the drop radius and the equilibrium contact angle. Working outside of this range can cause a switch in the direction of motion. This change can be added to (and interplay with) the switch in directionally given by the structure relative phase (between IP and OOP DOFs). For these reasons, to properly characterize the behavior of our systems, we should maintain the same drop modal response for all values of $f$. To this end, during the tracking of drop velocities, we adjust the drop size at each frequency (tables ~\ref{table:single-cell} and ~\ref{table:two-cell}) using the following equations:

\begin{equation}
    f_0^r = \frac{1}{2 \pi R^3} \sqrt{\frac{6 \gamma h(\theta)}{\rho (1-cos \theta)(2+cos \theta)}},
    \label{eq:rockingmode}
\end{equation}
\begin{equation}
    V = \frac{\pi R^3}{3} \frac{(1 - cos \theta)^2 (2 + cos \theta)}{sin^3 \theta},
    \label{eq:dropcapvolume}
\end{equation}

where $f_0^r$ is the first resonant frequency (\textit{rocking} mode), $R$ is the radius of the drop, $V$ is the volume, $\rho$ is the density, $h(\theta)$ is a numerical value that depends on the wetting angle, $\gamma$ is the surface tension between liquid and vapor, and $\theta_{eq}$ is the equilibrium contact angle. This expression (Eq.~\ref{eq:rockingmode}) is used along with the numerical values of $h(\theta)$ (also displayed in~\cite{Celestini2006}) to determine 
What drop size should we use (lower bound with the first rocking mode) to guarantee the same modal response for all frequencies with which we work in the single-cell and two-cell experiments. To find an estimate of the upper bound, we follow the discussion in~\cite{Noblin2009}, where they assume an approximation for the pumping mode as the $n=2$ Lamb's mode, therefore we estimate the upper bound as approximately twice the value of the lower bound. In Fig.~\ref{fig:dropsize_rockingfrequency} we plot the drop size by frequency $f$ (lower and upper bound) following these assumptions.

\subsection{Details on experiments: single and two cell}

Tables~\ref{table:single-cell} and~\ref{table:two-cell} contain information about drop size, frequency, and input amplitude (measured in voltages peak-to-peak and scaled through an amplifier model Br\"{u}el \& Kj\ae r Type 2718 Power Amplifier) for our two experiments: \textit{single-cell} and \textit{two-cell}, respectively. The cases with an asterisk (*) were subjected to a boosted input excitation to compensate for a lower displacement response (predicted and indicated by the spectral amplitude FE simulations in the main text).\\

The drop velocities were measured from the experiment videos. The frames where the drop started and stopped moving and the pixel location of the drop in each frame were recorded. The frame rate and scale (px/mm) were then used to calculate the drop velocity. 

\subsection{Two-cell alternative case}

\begin{figure}[t!]
\includegraphics[width=\columnwidth]{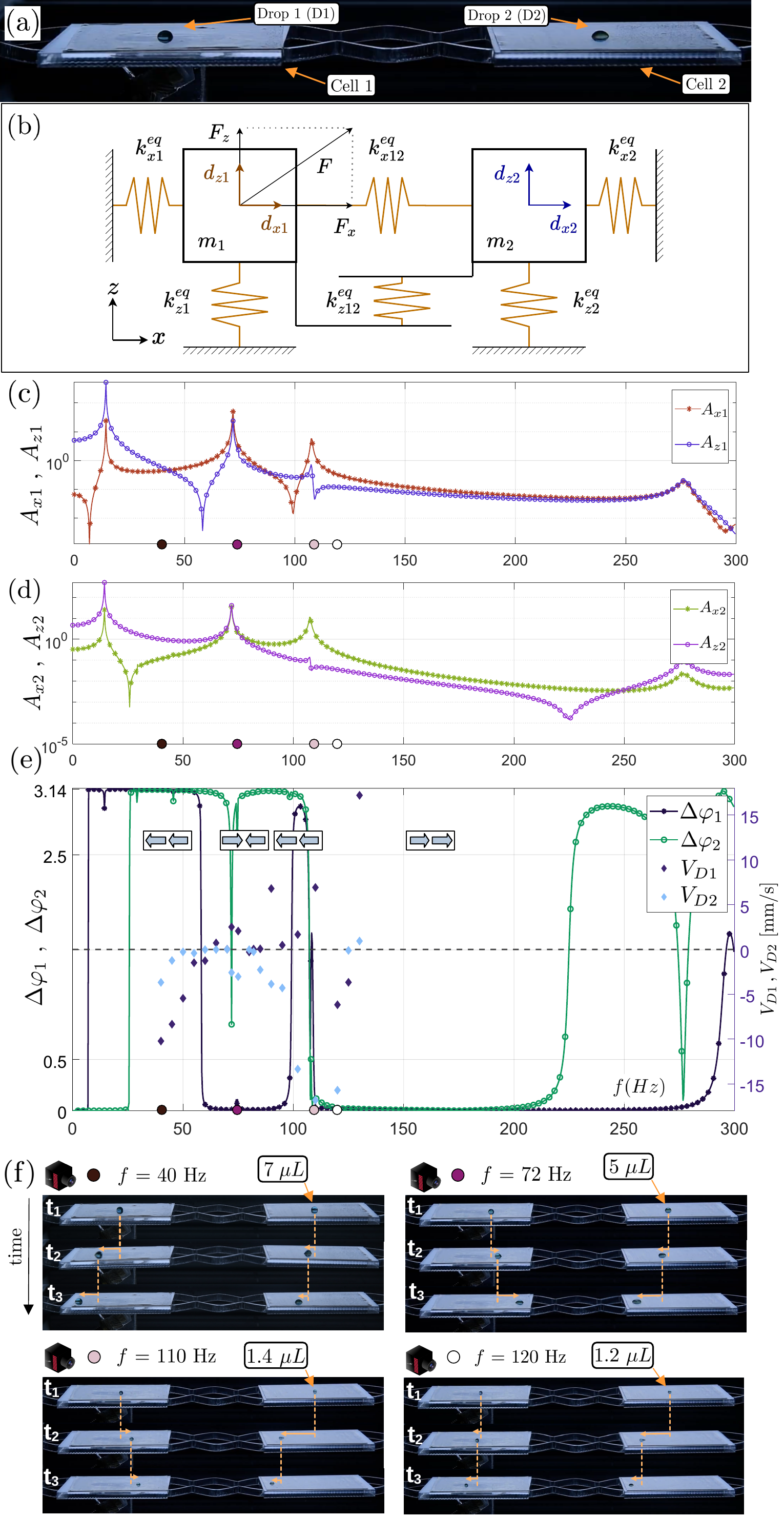}
\label{Alt_prototype}
\caption{\label{fig:panel3alt}(a) Two-cell prototype. (b) Lumped-parameter model with masses $m_1$ and $m_2$ and equivalent springs. 
A harmonic force $F$ is applied at cell 1. (c-d) Spectral amplitudes $A_{x1}$ (IP) and $A_{z1}$ (OOP) for cell 1 and $A_{x2}$ (IP) and $A_{z2}$ (OOP) for cell 2, from FE results. (e) Relative phases $\Delta \varphi_1$ and $\Delta \varphi_2$ determining the R vs. L motion of their respective drops. Average drop velocities experimentally acquired at sampled frequencies are overlaid to the plot, showing substantial correlation with the $\Delta \varphi$ curves. (f) High-speed camera snapshots of drop motion, revealing activation of LL, RL (convergent), LL, and RR regimes at frequencies $40 \, \textrm{Hz}$, $72 \, \textrm{Hz}$, $110 \, \textrm{Hz}$, $120 \, \textrm{Hz}$, and $130 \, \textrm{Hz}$.}
\end{figure}

We also present an alternative prototype for the two-cell system in Fig.~\ref{fig:panel3alt}, where the center springs are doubled in length (by placing two identical original springs in series). This modification creates a case where the equivalent stiffness of the inter-cell connection has half the stiffness of the external connections. The purpose of this design is to elicit the missing scenario where drops can move towards each other (RL or convergent, mentioned in the main text). The plotted data and snapshots in  Fig.~\ref{Alt_prototype} confirm the achievement of this additional regime that had eluded our initial design. 

\subsection{Tuned Cell Experiment}

\begin{figure}[t!]
\includegraphics[width=\columnwidth]{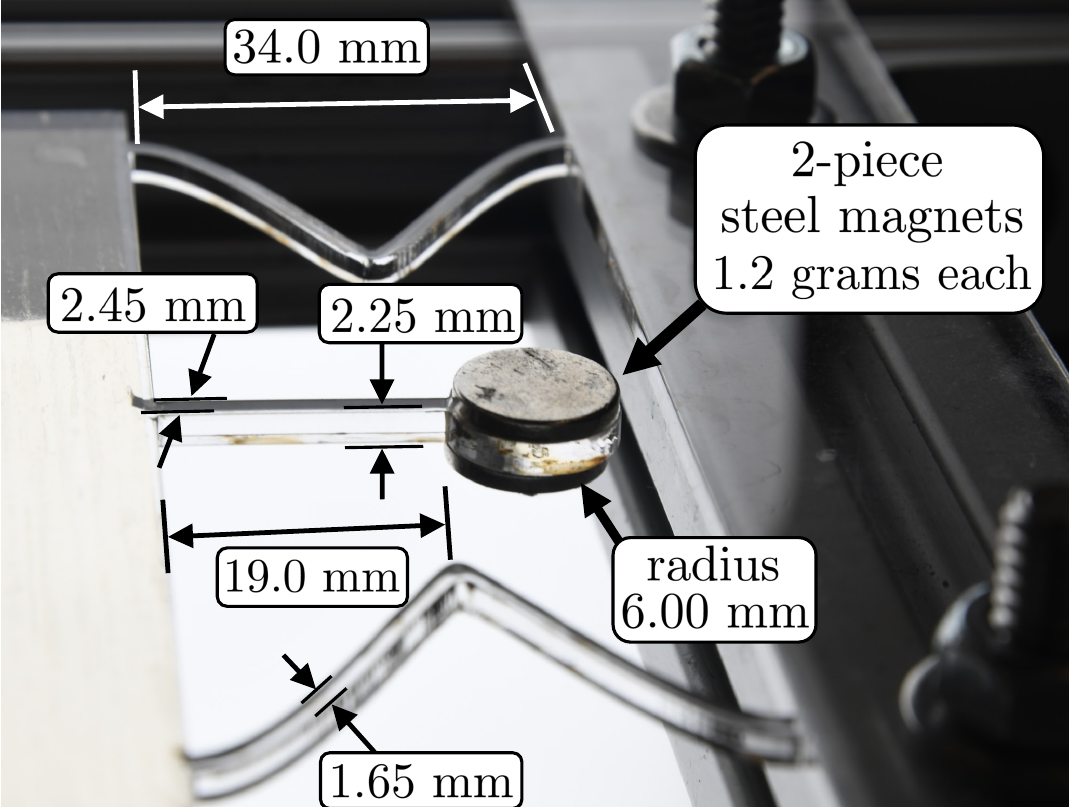}
\caption{\label{fig:resonators_detail} Modified cell details. Special note for the cell thickness, which in this case is 2.25 mm, differently from the 3.20 mm for the first single-cell prototype.}
\end{figure}

As mentioned in the main text, we build a modified version of the single-cell: 
details are shown in Fig.~\ref{fig:resonators_detail}.


\begin{table}[t!]
    \centering
    \begin{tabular}{|c|c|c|}
        \hline
        \textbf{Frequency (Hz)} & \textbf{Drop size ($\mu L$)} & \textbf{Velocity (mm/s)} \\
        \hline
        40 & 7.0 & 0.002 \\
        43 & 7.0 & 0.011 \\
        46 & 7.0 & 0.023 \\
        49 & 7.0 & 0.043 \\
        52 & 7.0 & 0.012 \\
        55 & 7.0 & 0.013 \\
        58 & 5.0 & 0.361 \\
        61 & 5.0 & 0.467 \\
        64 & 5.0 & 0.574 \\
        67 & 5.0 & 0.772 \\
        70 & 5.0 & 0.642 \\
        73 & 5.0 & 1.050 \\
        76 & 5.0 & 0.789 \\
        79 & 5.0 & 0.585 \\
        82 & 5.0 & 1.922 \\
        85 & 5.0 & 1.626 \\
        88 & 5.0 & 1.440 \\
        91 & 5.0 & 0.541 \\
        94 & 5.0 & 0.238 \\
        97 & 5.0 & 0.049 \\
        100 & 1.4 & 0.090 \\
        103 & 1.4 & -0.199 \\
        106 & 1.4 & -0.149 \\
        109 & 1.4 & -0.108 \\
        112 & 1.4 & -1.733 \\
        115 & 1.4 & -3.985 \\
        118 & 1.4 & -5.157 \\
        121 & 1.4 & -4.626\\
        124 & 1.4 & -4.579 \\
        127 & 1.4 & -2.182 \\
        130 & 1.4 & -0.051 \\
        133 & 1.4 & 0.384 \\
        \hline
        \hline
        130 & 1.2 & -4.562 * \\
        133 & 1.2 & 2.053 * \\
        136 & 1.2 & 1.681 * \\
        139 & 1.2 & 2.994 * \\
        142 & 1.2 & 0.857 * \\
        145 & 1.2 & 1.821 * \\
        148 & 1.2 & 1.468 * \\
        151 & 1.2 & 2.131 * \\
        154 & 1.2 & 2.077 * \\
        157 & 1.2 & 2.172 * \\
        160 & 1.2 & 1.632 * \\
        163 & 1.2 & 2.333 * \\
        166 & 1.2 & 1.368 * \\
        169 & 1.2 & 1.478 * \\
        172 & 1.2 & 1.472 * \\
        175 & 1.2 & 1.306 * \\
        178 & 1.2 & 0.430 * \\
        181 & 1.2 & 0.765 * \\
        \hline
    \end{tabular}
    \caption{Details on the \textit{single-cell} experiment. The cases with an asterisk (*) were subjected to a 40\% boosted input (from 2.5 volts to 3.5 volts).}
    \label{table:single-cell}
\end{table}

\begin{table}[t!]
    \centering
    \begin{tabular}{|c|c|c|}
        \hline
        \textbf{Frequency (Hz)} & \textbf{Drop size ($\mu L$)} & \textbf{Velocity D1\&D2 } \\
        \hline
        45 & 7.0 & -12.191 \& -13.129 \\
        50 & 7.0 & -5.698 \& -4.539 \\
        60 & 7.0 & -3.531 \& -1.778 \\
        65 & 6.0 & -3.696 \& -0.316 \\
        70 & 6.0 & -0.020 \& 0.007 \\
        75 & 5.0 & 0.000 \& 0.006 \\
        80 & 5.0 & 0.016 \& 0.102 \\
        85 & 5.0 & 0.027 \& 1.743 \\
        90 & 5.0 & 2.206 \& 3.292 \\
        95 & 5.0 & 3.764 \& 4.056 \\
        100 & 1.5 & 4.808 \& 11.718 \\
        105 & 1.5 & 0.853 \& 13.081 \\
        110 & 1.5 & 0.102 \& 19.219 \\
        115 & 1.2 & -1.520 \& 16.997\\
        120 & 1.2 & -0.379 \& 13.257 \\
        125 & 1.2 & -1.901 \& 11.208 \\
        130 & 1.2 & -0.038 \& 3.882 \\
        135 & 1.2 & -0.009 \& -2.422 \\
        140 & 1.2 & -1.379 \& -6.911 \\
        145 & 1.2 & 0.140\& -4.073 \\
        150 & 1.2 & 0.000 \& -2.856 \\
        155 & 1.2 & 0.000 \& -0.649 \\
        160 & 1.2 & 2.525 \& 1.087 \\
        165 & 1.2 & 3.435 \& 0.966 \\
        170 & 0.8 & 0.092 \& 0.504 \\
        \hline
        \hline
        175 & 0.8 & 4.683 \& 2.833 * \\
        180 & 0.8 & 7.544 \& 0.043 * \\
        185 & 0.8 & 20.849 \& 6.288 * \\
        \hline
    \end{tabular}
    \caption{Details on the \textit{two-cell} experiment. Velocities D1 and D2 of the drops placed on the cell 1 and cell 2 (mm/s), respectively. The cases with an asterisk (*) were subjected to a 90\% boosted input (from 2.2 volts to 4.2 volts).}
    \label{table:two-cell}
\end{table}

\begin{table}[h!]
    \centering
    \begin{tabular}{|c|c|c|}
        \hline
        \textbf{Frequency (Hz)} & \textbf{Drop size ($\mu L$)} & \textbf{Velocity D1\&D2} \\
        \hline
        40 & 7.0 & -10.226 \& -3.680 \\
        45 & 7.0 & -8.347 \& -1.236 \\
        50 & 7.0 & -5.454 \& -0.270 \\
        55 & 6.0 & -1.483 \& -0.413 \\
        60 & 6.0 & -1.260 \& -0.025 \\
        65 & 5.0 & 0.720 \& -0.015 \\
        70 & 5.0 & 0.026 \& 0.014 \\
        72 & 5.0 & 2.503 \& -2.582 \\
        75 & 5.0 & 2.053 \& -3.032 \\
        80 & 5.0 & -0.285 \& -0.030 \\
        82 & 1.5 & 0.043 \& -0.259 \\
        85 & 1.5 & 0.064 \& -2.270 \\
        90 & 1.5 & 6.795 \& -3.863 \\
        95 & 1.2 & 0.479 \&  -4.300 \\
        \hline
        \hline
        102 & 0.8 & 1.664 \& -13.349 * \\
        110 & 0.8 & 6.915 \& -16.815 * \\
        120 & 0.8 & -6.174 \& -15.709 * \\
        125 & 0.8 & -3.671 \& -0.114 * \\
        \hline
        \hline
        130 & 0.8 & 17.198 \& 0.940 ** \\
        \hline
    \end{tabular}
    \caption{Details on the \textit{two-cell} system with doubled springs experiment. Velocities D1 and D2 of the drops placed on the cell 1 and cell 2 (mm/s), respectively. The cases with an asterisk (*) were subjected to a 13\% boosted input (from 2.2 volts to 2.5 volts). The one with double asterisk (**) was subjected to a 240\% boosted input (from 2.5 volts to 6.0 volts).}
    \label{table:two-cell_double_springs}
\end{table}

\clearpage

\clearpage


\end{document}